\documentclass[10pt,letterpaper]{article}
\usepackage[top=0.85in,left=2.75in,footskip=0.75in,marginparwidth=2in]{geometry}

\usepackage[utf8]{inputenc}
\usepackage[T1]{fontenc}
\usepackage{lmodern}
\usepackage{cite}
\usepackage{float}
\usepackage{amsmath}
\usepackage{nameref,hyperref}

\usepackage{microtype}
\DisableLigatures[f]{encoding = *, family = * }

\raggedright
\usepackage{changepage}

\usepackage[aboveskip=1pt,labelfont=bf,labelsep=period,singlelinecheck=off]{caption}

\makeatletter
\renewcommand{\@biblabel}[1]{\quad#1.}
\makeatother

\usepackage{lastpage,fancyhdr,graphicx}
\usepackage{epstopdf}
\fancyheadoffset[L]{2.25in}
\fancyfootoffset[L]{2.25in}

\usepackage{color}

\definecolor{Gray}{gray}{.25}

\usepackage{graphicx}

\usepackage{sidecap}

\usepackage{wrapfig}
\usepackage[pscoord]{eso-pic}
\usepackage[fulladjust]{marginnote}
\reversemarginpar

\begin{document}
\vspace*{0.35in}

\begin{flushleft}
{\Large
\textbf\newline{Modeling Cancer Progression: An Integrated Workflow Extending Data-Driven Kinetic Models to Bio-Mechanical PDE Models}
}
\newline
\\
Navid Mohammad Mirzaei \textsuperscript{1},
Leili Shahriyari \textsuperscript{1,*}
\\
\bigskip
\bf{1} Department of Mathematics and Statistics, University of Massachusetts Amherst, Amherst, MA 01003, USA
\\
\bigskip
* lshahriyari@umass.edu

\end{flushleft}

\section*{Abstract}
Computational modeling of cancer can help unveil dynamics and interactions that are hard to replicate experimentally. Thanks to the advancement in cancer databases and data analysis technologies, these models have become more robust than ever. There are many mathematical models which investigate cancer through different approaches, from sub-cellular to tissue scale, and from treatment to diagnostic points of view. In this study, we lay out a step-by-step methodology for a data-driven mechanistic model of the tumor microenvironment. We discuss data acquisition strategies, data preparation, parameter estimation, and sensitivity analysis techniques. Furthermore, we propose a possible approach to extend mechanistic ODE models to PDE models coupled with mechanical growth. The workflow discussed in this article can help understand the complex temporal and spatial interactions between cells and cytokines in the tumor microenvironment and their effect on tumor growth.


\section*{Introduction}
Cancer is one of the main causes of death in today's world. The estimated number of new cancer patients in the United States for the year 2023 is 1,958,310, with an approximation of 609,820 deaths \cite{siegel2023cancer}. From 1971 to 2021, the federal investment in cancer research alone has increased from \$500 million to \$6.5 billion \cite{brawley202150}. There are many approaches to understanding cancer, and many interdisciplinary research groups have been formed to find a treatment for this disease. In more recent years, the use of mathematical tools to understand complex biological systems has become widespread, and cancer is not an exception.

Mathematical models, which have been used to understand complex mechanisms in cancer progression, can help us to create digital twins of cancer patients \cite{Stahlberg2022}. The non-spatial models are one popular approach to mathematically describe the interactions and dynamics of cancer. These models are time-dependent but do not consider spatial changes and interactions. Ever since Gompertz \cite{gompertz1825xxiv} proposed his simple model for cell population growth based on cell replication and death, many researchers have used similar or improved versions of this model to describe the dynamics of cancer progression.  This includes the use of logistic growth \cite{tsoularis2002analysis}, mass action laws \cite{horn1972general}, cooperative kinetics or Michaelis-Menten laws \cite{de1979kinetics}, and etc. For example, Kronik et al. use a system of 6 Ordinary Differential Equations (ODEs) describing Glioblastoma \cite{kronik2008improving}. In their model, they include the interaction between tumor cells, cytotoxic cells, cytokines such as TGF$\beta$ and IFN$\gamma$, and major histocompatibility complex class I and II. They use mass action, Michaelis-Menten and logistic growth laws in their model. They conclude that the alloreactive cytotoxic T-cell immunotherapy is a promising approach to controlling the tumor. Wang et al. take a Quantitative systems pharmacology (QSP) modeling approach to investigate the triple-negative breast cancer \cite{wang2022dynamics}. In these types of studies, the ODE system is broken into different modules, each having boundary variables that can connect to other modules. Using ten modules with a total of 150 ODEs, they capture the macrophage heterogeneity in the tumor microenvironment while maintaining its prediction accuracy at the population level. Sofia et al. use a similar mechanistic approach to model the Clear Cell Renal Cell Carcinoma (ccRCC) microenvironment \cite{sofia2022patient}. They show that IL-6 plays a significant role in controlling tumor cell populations. There are also stochastic models that can be classified as non-spatial models. 
Paterson et al. use a stochastic model of possible genotype mutations in colorectal cancer to quantify its risk of malignancy \cite{paterson2020mathematical}. 

Despite their usefulness, non-spatial models often describe a well-mixed population or culture. In other words, they neglect spatial heterogeneity. Also, they cannot capture the effect of a growing domain on the dynamics happening inside of it. Imagine a well-mixed population of sick and healthy people, and compare that with a population in which healthy people have less contact with sick people. Clearly, the outcome will be substantially different, and distinguishing these cases requires the inclusion of spatial dependence. There are many biological phenomena that can benefit from, including spatial dependence. Additionally, dynamics that lead to growth or shrinkage of their environment can be better explained this way. For example, Hao \& Friedman, use a system of PDEs to model the effect of LDL-HDL ratio on atherosclerotic plaque growth \cite{hao2014ldl}. They use a mechanistic interaction between the plaque constituents and model the growth using the conserved total density of cells and debris. They assume that surpassing this conserved quantity triggers plaque growth. Mohammad Mirzaei et al., study a similar problem \cite{mohammad2020integrated}, but they use the theory of morphoelasticity \cite{rodriguez1994stress} to model the growth. Similar approaches have been taken in the context of cancer modeling. Mohammad Mirzaei et al. extends ODE models of breast cancer and colorectal cancer to reaction-diffusion-advection models coupled with mechanical growth to investigate the effect of immune cells heterogeneity on cancer progression \cite{mohammad2022pde,mirzaei2023investigating}. Walker et al. use morphoelasticity in a simple model to investigate the effect of mechanics on tumor spheroid growth \cite{wang2022dynamics}. Mark AJ Chaplain's group, often uses a different method for a growing tumor domain. They consider a fixed frame for the tumor and the extracellular matrix hosting it. Then they assume a gradient of matrix-degrading enzymes secreted by cancer cells destroying the extracellular matrix and providing a channel for the tumor cells to diffuse \cite{domschke2014mathematical,trucu2013multiscale}.

Developing mathematical models to investigate a biological model is an intricate and challenging process. Coming up with a mathematical model is just the first step of the process. After that, one must estimate the parameters of the model. This requires access to data that is appropriately extracted for the desired biological system. Then the sensitivity of the model to the estimated parameters should be assessed to have an idea of the uncertainty of the results. Model extensions and assumptions should be considered carefully, and the results should be compared with the experiments. In this article, we provide a review of a workflow adopted by our lab. This workflow starts from mechanistic patient-specific ODE models of different tumor microenvironments \cite{le2021data,kirshtein2020data,sofia2022patient,mohammad2021mathematical,mohammad2022investigating} and extends to PDE systems coupled with mechanical growth \cite{mohammad2022pde,mirzaei2023investigating}. Various databases, parameter estimation, and sensitivity analysis techniques are practiced and explained in this article. 

\section*{A workflow of cancer modeling}
\begin{figure}[H]
    \centering
    \includegraphics[scale=0.6]{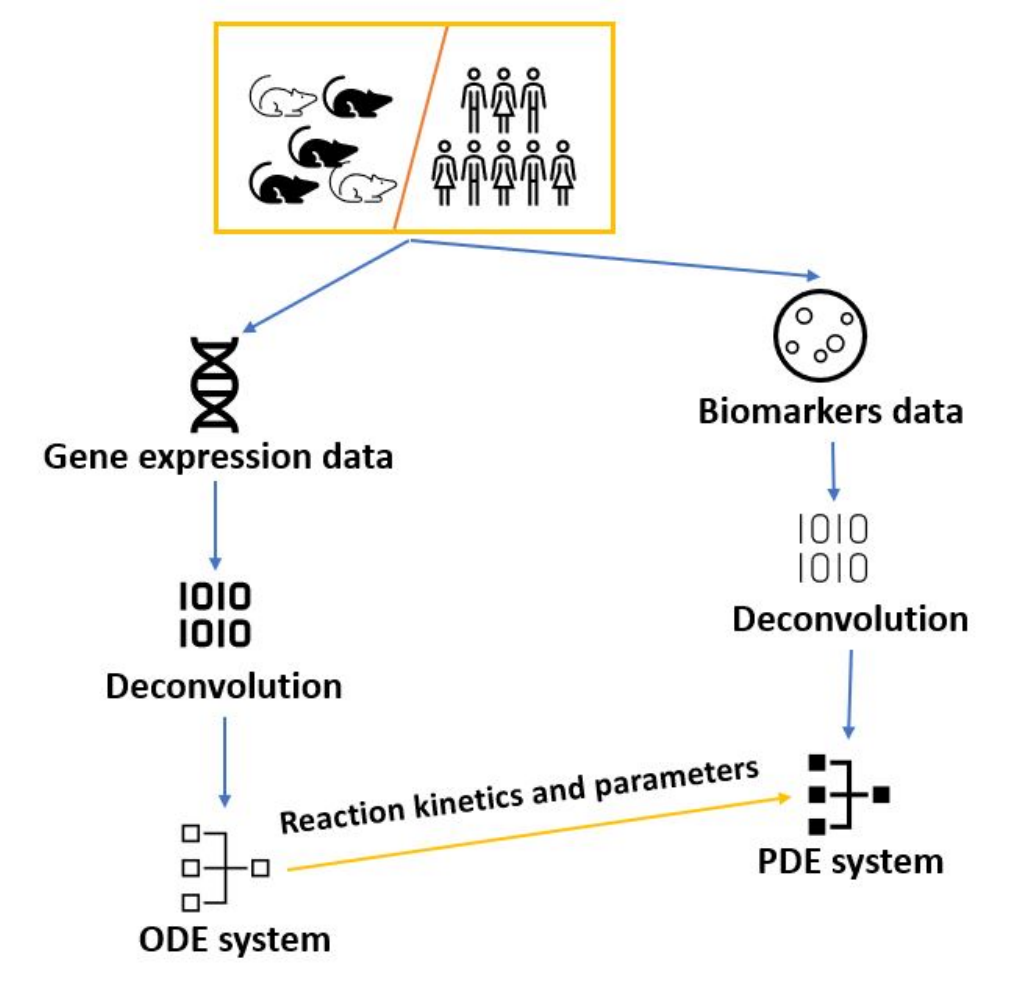}
    \caption{A schematic of the workflow discussed in this article.}
    \label{fig:workflow}
\end{figure}

We apply a systematic workflow to model different cancer types, which starts with acquiring and analyzing data. We then investigate the dynamics of key cells and molecules and their interactions in the tumor microenvironment. We create an ODE system that describes these interactions and estimates the kinetic parameters using the data. We study long-term behaviors and sensitive parameters. Having access to spatiotemporal data, we extend our model to a biomechanical PDE model. This allows us to study the spatial interactions and the effect of cell colocalization on the dynamics and spatial patterns of the microenvironment. Figure \ref{fig:workflow} shows a schematic of the workflow discussed here.
\subsection*{Acquiring data: Human vs. Mouse}
There are several private and public cancer data sets for human tumors. The Cancer Genome Atlas (TCGA) and The International Cancer Genome Consortium (ICGC) are the most comprehensive examples of public data sets. Initiated in 2006, TCGA provides data on gene expression, somatic mutations, copy number, single nucleotide polymorphisms (SNPs), microRNAs, clinical outcomes, and tissue slide images in addition to the clinical information of the patients \cite{cancer2013cancer}. ICGA was launched in 2008 and provides the same data types plus miRNA expression, DNA methylation, and exon junction \cite{international2010international}.
Other databases include other species, such as mice, rats, monkeys, and dogs. For example, Genevestigator is a public database containing gene expression data from all the aforementioned species besides humans \cite{zimmermann2004genevestigator}. Another public database covering multi-species data is the Gene Expression Omnibus (GEO) database \cite{edgar2002gene}. 

The benefit of mouse models is their abundance and flexibility. Extracting time course data from mice is more feasible than from humans. Often, the human gene expression data is acquired at first and a follow-up visit. Despite its abundance and flexibility, mice models might not fully represent human tumors precisely. In other words, human tumor microenvironments are usually more complex than their mouse counterparts. 

\subsection*{Data preparation}
We extract the percentage of immune cells in the tumor by applying tumor deconvolution methods \cite{Le2020,Aronow2022} to the tumor's gene expression data. CIBERSORTx is one of the most efficient deconvolution methods \cite{newman2019determining}. This step is applicable to both mouse and human data. 

Using human gene expression data, we usually have access to many patients with different tumor sizes. As mentioned before, one disadvantage of human data is the lack of time course information. However, after the deconvolution, we can group the patients into smaller clusters based on their immune profile using the K-means clustering. Mohammad Mirzaei et al. \cite{mohammad2021mathematical} apply this procedure on 2993 breast cancer patient data acquired from TCGA and the Molecular Taxonomy of Breast Cancer International Consortium (METABRIC) cohort. They end up with 5 clusters of patients with similar immune profiles. 

We can access the time course data of mouse models much easier. There are usually a few mice in the same study. However, each mouse tumor microenvironment can be sampled several times before sacrifice. Due to a lack of complexity, Mouse models are carefully picked to mimic human tumors as closely as possible. MMTV-PyMT mouse models are very popular in breast cancer studies. These models are engineered to show the signaling of receptor tyrosine kinases, which are activated in many human progressive breast tumors \cite{attalla2021insights}. Mohammad Mirzaei et al. \cite{mohammad2022investigating} use data from three MMTV-PyMT mice from Gene Expression Omnibus (GEO) database \cite{cai2017transcriptomic}. The gene expression data were acquired at four time points: hyperplasia at week 6, adenoma/MIN at week 8, early carcinoma at week 10, and late carcinoma at week 12. Using gene expression deconvolution at each time point, they arrived at the immune cell percentages in the tumor microenvironment.

For both data types (human and mouse), the amount of cancer and necrotic cells was estimated using cancer:necrotic:immune cell ratios reported in the literature for each species. The immune cells, cancer, and necrotic frequencies were later used for initial condition and parameter estimation purposes.

\subsection*{The interaction network and ODE model}
There are numerous biological studies focused on the effect of cells and cytokines on each other in the tumor microenvironment. These studies provide the groundwork for building an interaction network that inspires the ODE model. 

One way of mathematically interpreting these networks is to use the law of mass action. Though often used for chemical reactions, this law can be used in linear form to model positive and negative feedback loops. For example, if in a hypothetical situation cell or cytokine $A$ reacting with cell or cytokine $B$ leads to an increase in the levels of $B$, at rate $\lambda_{AB}$, we say:
\begin{equation}
    \frac{dB}{dt}= \lambda_{AB} A B. \label{eq:posfeedback}
\end{equation}
Conversely, if interaction with cell or cytokine $C$ leads to a decrease in the levels of $B$, at rate $\delta_{CB}$, we write:
\begin{equation}
    \frac{dB}{dt}= -\delta_{CB} C B. \label{eq:negfeedback}
\end{equation}
Some researchers also use other methods, such as Michaelis-Menten kinetics. It is worth mentioning that extra caution is needed for such methods since reversibility and reactant-stationary assumptions during the initial transient period are required \cite{schnell2014validity,stroberg2016estimation}. For example, in \cite{mohammad2021mathematical,mohammad2022investigating,kirshtein2020data}, the authors have assumed that the cell-cell interaction of cytotoxic T-cells and cancer cells is enough to lead to cancer cell inhibition. While in \cite{sofia2022patient}, an intermediate kinetic is involved; i.e., the inhibition occurs only when PD-1 and PD-2 proteins expressed by cytotoxic T-cells attach to the ligands PD-L1 and PD-L2 expressed by cancer cells. This leads to Michaelis-Menten-like inhibition rates. 

Moreover, we can avoid blow-ups by using nonlinearity in our ODEs. For example, using a logistic growth model for cells that do not differentiate from other cells, such as cancer and adipocytes. As for cells that do differentiate/activate from a naive state, there is no need for increased complexity since the initial population of the naive cells will prevent the blow-up of the resulting activated cells. Helper, cytotoxic and regulatory T-cells differentiate from naive T-cells, and activated dendritic and macrophages also differentiate from their naive cell types. The mentioned framework can be seen across all our lab publications \cite{mohammad2021mathematical,mohammad2022investigating,kirshtein2020data,le2021data,sofia2022patient}. 

\begin{figure}[H]
    \centering
    \includegraphics[scale=0.5]{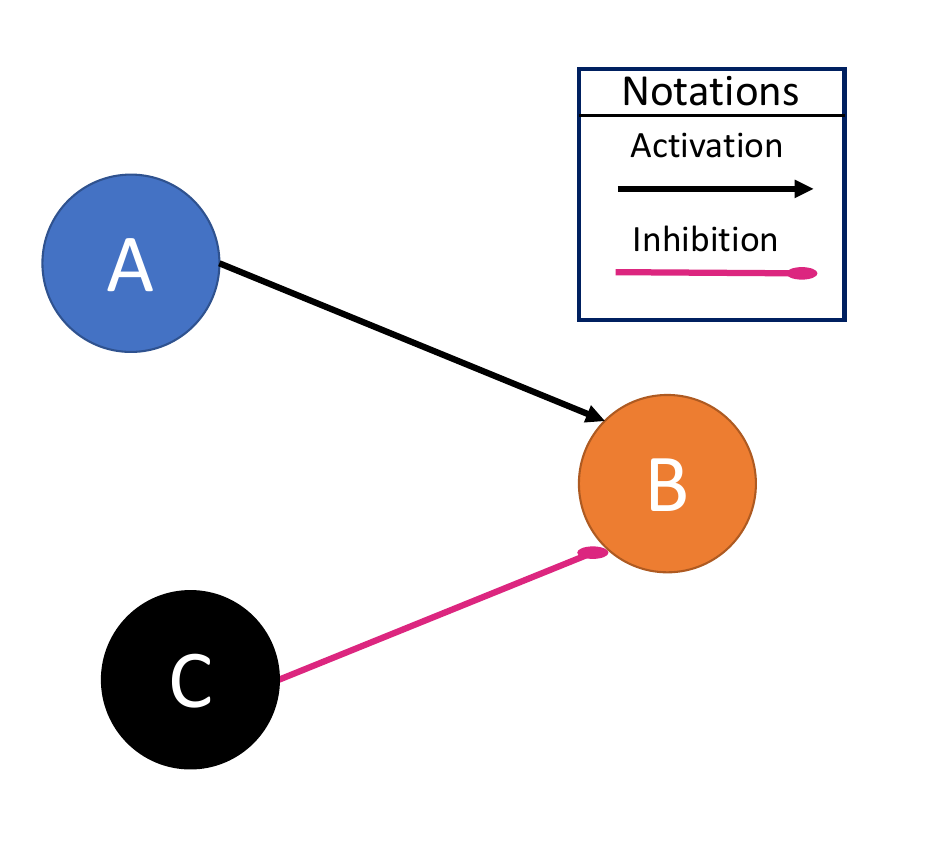}
    \caption{The interaction network corresponding to equation \eqref{eq:ODE1}. Intrinsic proliferation and natural death/decay are usually excluded from the graphical representations.}
    \label{fig:network}
\end{figure}

There are other terms that might be included depending on the nature of the cells and cytokines. Terms such as the intrinsic cell proliferation rate. This corresponds to proliferation which is independent of other cell-cell or cytokine-cell effects.  For a cell $B$, these proliferation and natural death rates can be modeled using the following, respectively:
\begin{eqnarray}
    \frac{d B}{dt} &=& \lambda_B B, \\
    \frac{dB}{dt} &=& -\delta_B B.
\end{eqnarray}
So overall, the ODE that governs the dynamics of a cell type $B$, which gets promoted by $A$ and inhibited by $C$ and has intrinsic proliferation and death, is given by:
\begin{equation}
    \frac{d B}{dt} = (\lambda_B+\lambda_{AB}A)B - (\delta_{B}+\delta_{CB}C)B \label{eq:ODE1}
\end{equation}
A graphical interaction network for this hypothetical situation is given in Figure \ref{fig:network}. 

Following this approach we get a system of ODEs for all key cells and cytokines in the tumor microenvironment. Namely,
\begin{equation}
    \frac{d [X_i]}{dt} = {f_i}({\bf [X]},\boldsymbol{\theta}) \label{eq:ODEsys}
\end{equation}
where ${\bf [X]}=\langle [X_1],\cdots,[X_M] \rangle$ and $\boldsymbol{\theta}= \langle \theta_1, \cdots, \theta_N \rangle$ are the vectors of state variables and parameters, respectively.
\subsection*{Parameter estimation}
There are several methods of parameter estimation. Which method to use is based on the data available. Overall, we can divide these methods into stochastic and deterministic. Each method has its advantages and disadvantages. Markov Chain Monte Carlo (MCMC) algorithms are among the most popular algorithms for stochastic parameter estimation. In these methods, a rough estimate of the parameters is enough, and it is applicable even without any experimental data. Even though access to the experimental data can improve the estimation, all that is really needed is a prior distribution for the parameter space. The disadvantage of these methods is that in big systems with many unknown parameters, they can be very slow. Also, the prior distribution is not always known. For more information on different MCMC approaches, see \cite{valderrama2019mcmc}. 

As for the deterministic method, access to the data is essential. Authors in \cite{le2021data,mohammad2021mathematical,sofia2022patient,kirshtein2020data}, use steady-state parameter estimation. This method only requires the value of the state variables at the steady state (i.e., $\frac{dx}{dt}=0$). For instance, if an ODE system consists of N variables and N unknown parameters, replacing the value of the state variables with their steady-state values will give a closed set of homogeneous algebraic equations that can be quickly solved for the parameter values. However, usually, the number of parameters is larger than the number of unknowns. Then more equations are needed to close the system. This is the main disadvantage of this method, but making biological assumptions about the parameters based on the literature can remedy this. For example, if we know the rate of cancer inhibition via cytotoxic T-cells is higher than IFN-$\gamma$, we can interpret it into a mathematical equation with the two non-dimensional rates being multiple of each other. Later, this assumption and the scaling need to be investigated for robustness by applying proper perturbations and investigating the effect on the overall system behaviors.

The other deterministic method which is commonly used is the least square optimization. This method is more helpful when we have access to time-course data. We adopted the least square method in \cite{mohammad2022investigating} due to access to time-course mice data. The goal is to minimize the square distance between the observed data and the predicted dynamics. In other words, if matrix $A$ has the observation at time $i$ for the state variable $j$ as its $ij$th entry, and the vector ${\bf b}$ contains the ODE predicted value at time $i$ in its $i$-th entry, then the least square optimization is given by:
\begin{eqnarray*}
&\text{Find}& \boldsymbol{\theta}:= \langle \theta_1, \cdots, \theta_N \rangle \ \text{such that} \\
&\underset{\boldsymbol{\theta}}{Min}& \frac{1}{2}|| A \boldsymbol{\theta} - {\bf b} ||_2^2 
\end{eqnarray*}
where $\boldsymbol{\theta}$ is the vector of unknown parameters. The disadvantage of this method is that it is highly data-dependent. The low number of data points and high disparity in the values of a few state variables can highly affect the method's precision. Including higher weights for more important measurements or adding regularization terms to the minimizer are ways to improve the results. For a quick survey on the parameter estimation methods, see \cite{mitra2019parameter}.

\subsection*{Global sensitivity analysis}
Sensitivity analysis is to quantify the sensitivity of the model to the parameters of the model. In this section, we cover the global sensitivity analysis, which is an appropriate method, especially when the parameters are estimated using steady-state assumptions. 
For an ODE given by 
\begin{equation}
\frac{dX}{dt}=f(X,\boldsymbol{\theta}), \label{eq:ODE}
\end{equation}
where $\boldsymbol{\theta}:= \langle \theta_1, \cdots, \theta_N \rangle$
the mathematical definition of the sensitivity of a state variable $X$ to a parameter $\theta_i$ is given by:
\begin{equation}
    s_i = \frac{dX}{d \theta_i}. \label{eq:sens1}
\end{equation}
This can be calculated in time or at a steady state. Assuming $\hat{X}$ is the value of the state variable at the steady-state, then equation \eqref{eq:ODE} turns into $f(\hat{X},\boldsymbol{\theta}) = 0$. Differentiating this equation with respect to $\theta_i$ by way of chain rule we get:
\begin{equation}
    \frac{\partial f}{\partial \hat{X}}\frac{d\hat{X}}{d \theta_i}+\frac{\partial f}{\partial \theta_i} = 0. \label{eq:sens2}
\end{equation}
The term $\frac{\partial f}{\partial \hat{X}}$ is non-singular, as long as our function $f$ is continuously differentiable. So it can be inverted, and thus equation \eqref{eq:sens2}, can be solved for the sensitivity $\frac{d\hat{X}}{d \theta_i}$ at the steady-state. 

To further improve this method we define the local sensitivity $S_i$ as:
\begin{equation}
    S_i = \displaystyle \int_{\Omega} s_i (\boldsymbol{\theta}) d\boldsymbol{\theta}, \label{eq:localsens} 
\end{equation}
where $\Omega$ is a chosen neighborhood of the base parameter set. When doing steady-state parameter estimation, the parameter values are acquired from solving a system of algebraic equations. Therefore, perturbing one will affect the rest, and \eqref{eq:localsens} obviates that dependence locally. A powerful numerical method to calculate the integral in \eqref{eq:localsens}, is using sparse grid points \cite{gerstner1998numerical}. 

Finally, we need to assess the sensitivity of our model to the extra assumptions we added to close the system of algebraic equations for steady-state parameter estimation. We can do this by calculating a waited average of local sensitivities attained by factoring the extra assumptions. For example, assume $p= 0.01, \cdots, 100$ are $N$ scaling factors. Then multiplying each assumption by a $p_k$ will create a new set of base parameters for which we can calculate the local sensitivity for $\theta_i$, namely $S_i^{p_k}$. Then, obviously, $S_i^1$ corresponds to no assumption scaling and gives sensitivity for $\theta_i$, for the original set of parameters. Then, the global sensitivity is defined by:
\begin{equation}
    \mathbf{GS}_i = \displaystyle \sum_{k=1}^N w_k S_i^{p_k}. \label{eq:gs}
\end{equation}
The weights $w_k$ should be picked so that the parameter values are closer to the original values and take larger weights than the ones that are very different from the base values. For more detailed information about the weights, see \cite{zi2011sensitivity}.

This sensitivity analysis has two important implications. Firstly, it introduces a set of parameters that allow us to investigate the robustness of the model. By perturbing this set of parameters and observing the dynamics we can attain an interval of confidence for our predictions. Secondly, it tells us what biological factors control important dynamics, such as the population of cancer cells. These parameters can later be targeted in a therapy model for controlling the cancer progression. 

\subsection*{ODE to PDE extension}
As mentioned in the introduction, several methods exist to model cancer growth. In this section, we cover the method that uses a conserved quantity to connect the biology of cancer to its mechanical growth. Mohammad Mirzaei et al. have used this approach to model breast and colorectal cancer coupled with their corresponding mechanical models, see \cite{mohammad2022pde,mirzaei2023investigating}. Hu et al. use a similar approach coupled with a porous medium mechanical model for osteosarcoma \cite{hu2022bio}.  

\begin{figure}[H]
    \centering
    \includegraphics[width=\textwidth]{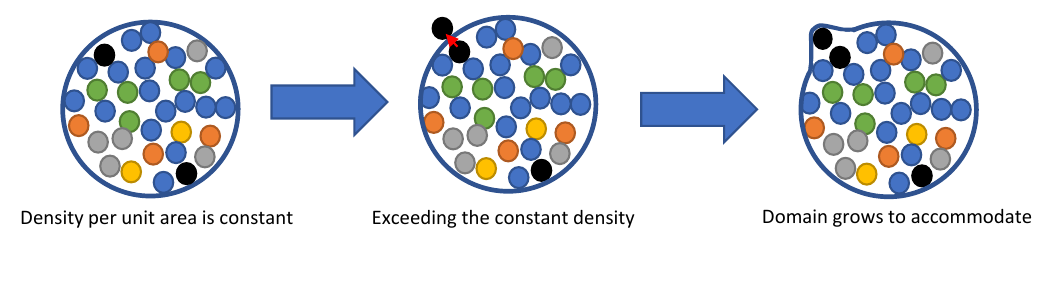}
    \caption{Schematic of growth based on conserved total density of cells.}
    \label{fig:mech}
\end{figure}

The backbone of this method is the assumption that the total density of cells per unit area is constant. This requires us to further assume that the cells are densely packed per unit area, and there is no room for adding any more. Therefore, when cells proliferate, they need to displace to another location which defines the domain movement velocity, see Figure \ref{fig:mech}

Now we go over the process of extending the ODE system to a PDE system and coupling it with mechanical growth. The spatial dependence introduced in PDEs provides an opportunity to model the movement of cells and cytokines in their environment. These movements can happen for several reasons. One can be a random and independent movement that can be modeled via a diffusion term. We can also include an advection of cells according to the overall movement of the domain. There are other reasons, such as chemotaxis, haptotaxis, and durotaxis which correspond to the movement of cells up or down a gradient of substances, substrate-bound molecules, or mechanical stress. We refer the reader to some articles which model these events \cite{hughes2012quantitative,painter2019mathematical,shangerganesh2019finite,wang2020review,rens2020cell}. However, to be able to decouple the biological model from the mechanical model using the conserved density method, we avoided these. Therefore, the system \eqref{eq:ODEsys} would be extended to the following reaction-diffusion-advection system:
\begin{equation}
    \frac{\partial [X_i]}{\partial t}+ b_i \nabla \cdot ({\bf v} [X_i]) = D_i \Delta [X_i]+f_i({\bf [X]},\boldsymbol{\theta}). \label{eq:PDE}
\end{equation}
where $D_i$ is the diffusion coefficient, ${\bf v}$ is the domain movement velocity, and $b_i=1$ for cells and $b_i=0$ for molecules. This is because molecules are orders of magnitude smaller than cells. Therefore, the domain deformation does not affect them, and they cannot significantly impact the domain deformation. Now, going back to the assumption that the total density of cells is constant, we can sum up the PDEs for the cells and derive a formula for the velocity:
\begin{equation}
    \nabla \cdot {\bf v} = \frac{\displaystyle\sum_{cells} f_i({\bf [X]},\boldsymbol{\theta})}{c}. \label{eq:divv}
\end{equation}
The subscript $cells$ means we are only summing over the right-hand sides corresponding to cell types. The denominator is the total density of cells which we assumed to be a constant. Equation \eqref{eq:divv} provides the connection between mechanical deformation/growth and biology. Now taking $Q({\bf v},p)$ to be a stress tensor appropriately chosen for the tumor type, then the balance of momentum and neglecting the inertia and body forces gives:
\begin{equation}
    \nabla \cdot Q = 0. \label{eq:stress}
\end{equation}
which we should solve for the velocity ${\bf v}$ and the pressure $p$. However, equation \eqref{eq:stress} by itself is underdetermined. This is when the biology-driven velocity from equation \eqref{eq:divv} comes into play and closes the system alongside equation \eqref{eq:stress}. As mentioned before, based on the type of tumor, we choose the stress tensor $Q$ to reflect the appropriate mechanical model. For example, for breast cancer, we can use a viscous low-velocity Stokes flow, or for colorectal cancer \cite{mohammad2022pde}, we can use a simple incompressible linear elasticity model \cite{mirzaei2023investigating}, or for osteosarcoma, a porous medium model \cite{hu2022bio}. 

\begin{figure}
    \centering
    \includegraphics[scale=0.5]{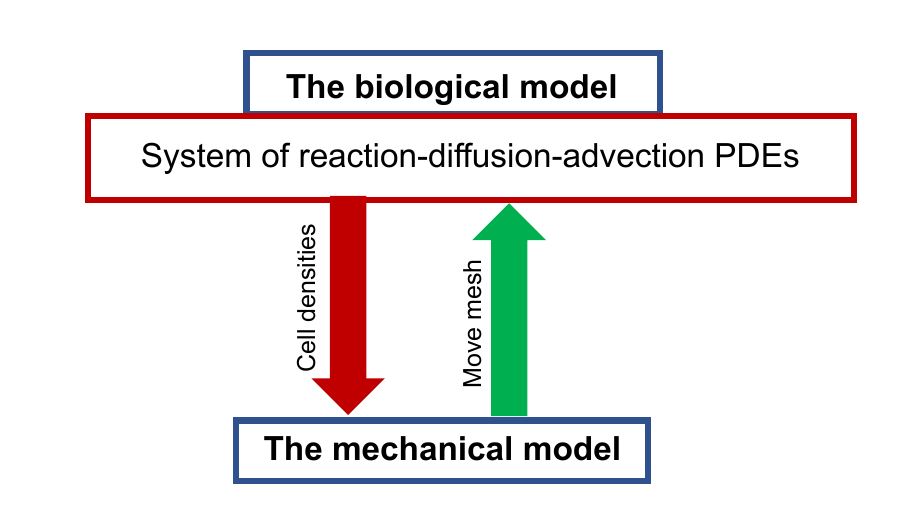}
    \caption{The process of solving the PDE system coupled with the mechanical model. We start by solving the biological PDE system for cell and cytokine densities. The cell densities will be the input and the trigger for the mechanical deformation. Once we get the velocity/displacement from the mechanical model, we use it to move the domain. In the next time step, we solve the PDE system in the deformed domain and use its results to solve the mechanical problem and so on. This continues until the last time step is reached.}
    \label{fig:process}
\end{figure}

Finally, for the boundary conditions, we can consider a free traction boundary condition, i.e., $Q{\bf n} = 0$, where ${\bf n}$ is the unit outward normal vector for the boundary. We can also use the boundary condition
\begin{equation}
    Q {\bf n} = - \gamma \kappa {\bf n} \label{eq:BCmech}
\end{equation}
where $\kappa$ is the mean curvature of the boundary and $\gamma$ is the cell-cell adhesion forces. This boundary condition is proposed by Franks et al. \cite{franks2003modelling}, and it helps prevent the boundary of the tumor from falling apart. Finally, we need to point out that these types of mechanical problems do not have a guaranteed unique solution. Since any deformed configuration can be rigidly translated and rotated and still be the solution. To remedy this, we can exclude these behaviors by enforcing the following constraints:
\begin{equation}
    \int_\Omega {\bf v} dx=0, \quad \int_\Omega {\bf v}\times {\bf x} dx=0.
\end{equation}
The first integral is for excluding rigid translations, and the second one is for excluding rigid body rotations. Figure \ref{fig:process} shows the process of solving the system of PDEs coupled with the mechanical deformation model.

\subsection*{Spatial data preparation}
In this section, we cover spatial data preparation for a system of PDEs based on spatial proteomics. This procedure was used by Mohammad Mirzaei et al. \cite{mohammad2022pde}.

To acquire the data, the tumor samples were stained with F4/80 and CSF1R antibodies and hematoxylin. Then using an iterative method including, staining, whole slide scanning, and antibody stripping. This method performed with multiplex immunohistochemistry (mIHC) can detect up to 6-30 biomarkers \cite{lewis2021spatial}. The biomarker signals were visualized using fluorescents such as anti-rabbit or anti-rat Histofine Simple Stain MAX PO
horseradish peroxidase (HRP)-conjugated polymer. The data was then processed using MATLAB's detectSURFFeatures algorithm from the Computer Vision Toolbox. According to this data, we can find out how much of a certain biomarker is expressed at each coordinate $(x,y)$ of the sample. 

\begin{figure}[H]
    \centering
    \includegraphics[width=\textwidth]{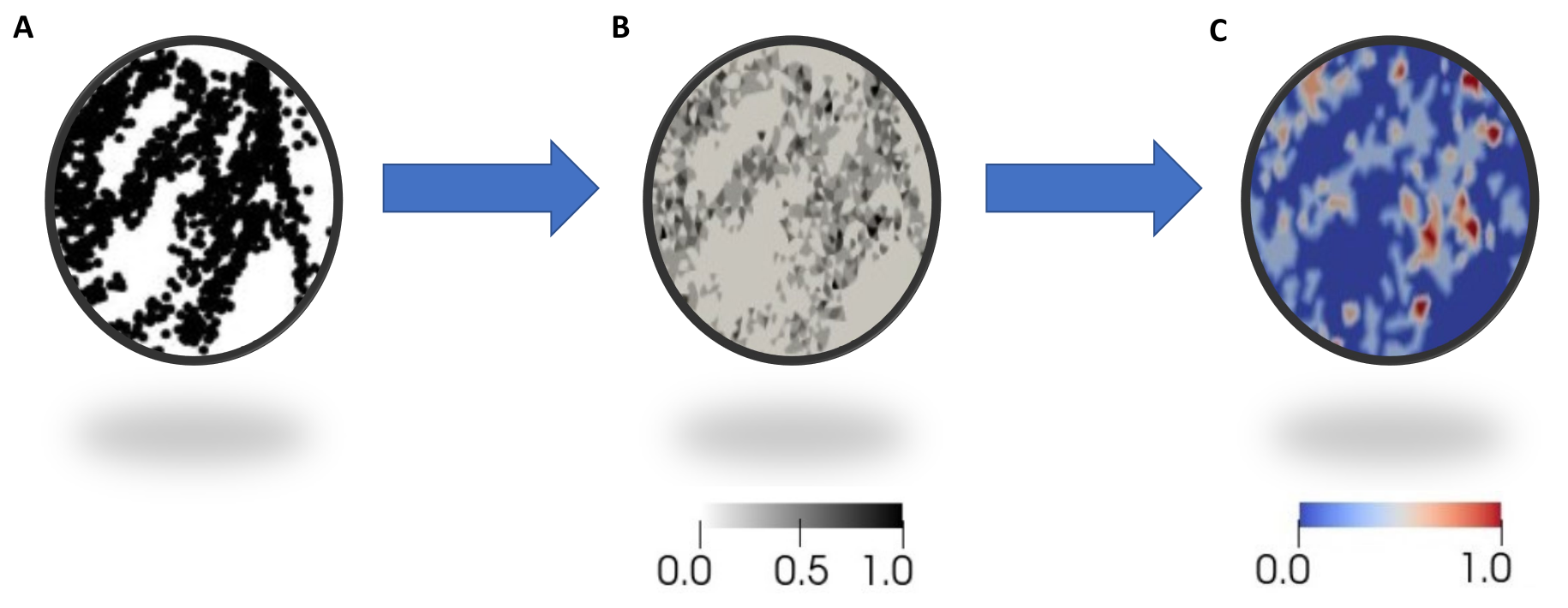}
    \caption{PDE initial condition preparation steps. (A) Cell locations are determined by using biomarker combinations. (B) A discontinuous Galerkin function is obtained by counting the number of cells from part (A) in each triangle and then normalizing. (C) Projecting the discontinuous Galerkin function from part (B) onto a continuous first-order Galerkin space creates the IC used in the PDE system.  }
    \label{fig:IC}
\end{figure}

We determine the presence and absence of each cell type in the initial tumor based on the presence and absence of a combination of biomarkers at a certain location. For example, if at a point the biomarkers CD45, CD3, and CD4 are expressed, but Epithelial Cell Adhesion Molecule (EpCAM) and CD8 are not (in short, EpCAM(-)CD45(+)CD3(+)CD4(+)CD8(-)), then there is a helper T-cell there. There are combination biomarkers for each cell type, and applying the same method for each will create a discontinuous field corresponding to the initial condition of each state variable, see Figure \ref{fig:IC}A.  
Now to a triangulated domain representing a discretized version of the tumor microenvironment, we assign a discontinuous Galerkin space of degree zero. Functions belonging to this space attain a constant value in each triangle. We then count the number of cells in each triangle and create a discontinuous Galerkin function as follows:
\begin{eqnarray}
    [X]({\bf x})  = \displaystyle \sum_{j=1}^m \omega_i(\mathcal{T}_j) \xi_j({\bf x}), \qquad \text{for} \quad i \in I, \label{eq:DG_IC}
\end{eqnarray}
where $\xi_i(\mathcal{T}_j)$ is the number of the cell type $[X]$ in the triangle $\mathcal{T}_j$. The function $\chi_j({\bf x})$ for a mesh point ${\bf x}=(x,y)$ is a characteristic function given by:

\[\chi_j({\bf x}) = \begin{cases}
 1 \qquad \text{if} \quad {\bf x} \in \mathcal{T}_j, \\
 0 \qquad \text{Otherwise}.
\end{cases}
\]
A sample visualization of these types of functions is given in Figure \ref{fig:IC}B.  
As mentioned in the previous section, our model is a continuum model in both biology and mechanics. This means discontinuous functions such as equation \eqref{eq:DG_IC} need to be approximated by continuous functions. For this purpose, we construct a linear continuous Galerkin function space. This space consists of linear hat functions on each triangle. Next, we project the functions of type \eqref{eq:DG_IC} onto this space. The result will resemble Figure \ref{fig:IC}C. Following the same procedure for each cell type, we will have a set of spatial initial conditions for our PDE system.

\subsection*{Adjoint-based sensitivity analysis}
Sensitivity analysis for PDEs can be trickier due to the sheer size of the solutions. Remember, that the solutions are now both time and space dependent and, for a fairly fine mesh, the derivative of the system with respect to the solution in \eqref{eq:sens2} entails a time-consuming process. In the global sensitivity analysis for the ODE system we discussed earlier, this task was less of a concern, given the short amount of time required for solving the system. This can be repeatedly performed for each parameter sensitivity, and for a system of 15 variables and 57 parameters, the whole task takes about 4-7 hours on a personal computer. In contrast, for a time and space dependent PDE system, it takes double that time to even get to the system solution, so carrying that out several times for each parameter is not feasible. 

Adjoint-based Sensitivity Analysis (ABSA), is a remedy for this issue \cite{li2004adjoint}. Take the following system of PDEs:
\begin{equation}
    {\bf F}({\bf [X]},\boldsymbol{\theta}) = 0 \label{eq:PDEsys}
\end{equation}
where ${\bf [X]}$ is implicitly dependent on parameters $\boldsymbol{\theta}$. Define $J({\bf [X]})$ to be the quantity of interest we want to measure its sensitivity to parameters, meaning $\frac{\partial J}{\partial \boldsymbol{\theta} }$. As mentioned, for a large system, the direct method is not computationally feasible. 

 The idea of ABSA is to take away the dependence of $\frac{\partial J}{\partial \boldsymbol{\theta} }$ on ${\bf [X]}$. From equation \eqref{eq:PDEsys}, we can conclude
\begin{equation}
    J({\bf [X]}) = J({\bf [X]}) +  \boldsymbol{ \epsilon}^T {\bf F}({\bf [X]},\boldsymbol{\theta}), \label{eq:19}
\end{equation}
holds for all vectors $\boldsymbol{ \epsilon}$ with proper dimension. Now, perturbing $\boldsymbol{\theta}$, will cause a  perturbation in ${\bf [X]}$ and $J$, consequently. So using a variational chain rule
By the way of {the} chain rule on \eqref{eq:19} we get:
\begin{equation}
     \delta J = \frac{\partial J}{\partial {\bf [X]}} \delta {\bf [X]} + \boldsymbol{ \epsilon}^T \left( \frac{\partial {\bf F}}{\partial {\bf [X]}} \delta {\bf [X]} + \frac{\partial {\bf F}}{\partial \boldsymbol{\theta}} \delta \boldsymbol{\theta} \right) =\underbrace{ \left( \frac{\partial J}{\partial {\bf [X]}}  +  \boldsymbol{ \epsilon}^T  \frac{\partial {\bf F}}{\partial {\bf [X]}} \right) \delta {\bf [X]}}_\text{(*)} + \boldsymbol{ \epsilon}^T \frac{\partial {\bf F}}{\partial \boldsymbol{\theta}} \delta \boldsymbol{\theta}. \label{eq:20}
\end{equation}
So if (*) equals zero, then we have achieved our goal of obviating the dependence on the solution ${\bf [X]}$. The equation 
\begin{equation}
    \left( \frac{\partial {\bf F}}{\partial {\bf {\bf [X]}}} \right)^T \boldsymbol{ \epsilon} = - \left( \frac{\partial J}{\partial {\bf {\bf [X]}}} \right )^T \label{eq:21}
\end{equation}
is known as the adjoint equation. Similar to the global sensitivity, since $\frac{\partial {\bf F}}{\partial {\bf [X]}}$ is non-singular by the implicit function theorem we are guaranteed to get a vector $\boldsymbol{\epsilon}$ which makes (*) in equation \eqref{eq:20} zero. Therefore, we can rewrite equation \eqref{eq:20} independent of ${\bf [X]}$:
\begin{equation}
\frac{\partial J}{\partial \boldsymbol{\theta}}  = \boldsymbol{\epsilon}^T \frac{\partial {\bf F}}{\partial \boldsymbol{\theta}}. \label{eq:22}
\end{equation}
Even though, we still need to calculate the solution ${\bf [X]}$ for a nominal set of parameters $\boldsymbol{\theta}$ from equation \eqref{eq:PDEsys} to get $\boldsymbol{\epsilon}$  from \eqref{eq:21}. But, this is only once, and not for every parameter sensitivity calculation since after that, the sensitivity calculation is independent of the solution.

\section*{Discussion}
In this review article, we proposed a possible methodology for modeling tumor microenvironments. The steps introduced here provide an opportunity for researchers to investigate the complex interactions occurring within the environment both from a dynamics point of view and spatial interactions. Also, such models can be used in a modular manner. For example, if we want to add vasculature to the model, we can design a separate system that models the vessel formation and then recognize the components of the vasculature model which interact with the components of the microenvironment model. These will be the boundary variables that enable us to join the two models. This will also be easier in terms of parameter estimations and model maintenance.   

We discussed some available databases and the advantages and disadvantages of using human vs. mouse models. Though human data is accessible through large public databases, time-course data is rare. On the contrary, mice models are more flexible, and acquiring time course data from them is more feasible. However, they lack complexity and sometimes do not fully reflect some important human tumor components. We offered data preparation ideas that can circumvent some of these shortcomings. For example, clustering human cancer patients based on the similarity of their tumor immune profiles can mimic a patient with different tumor stages. In other words, within each cluster, we can sort patients by their tumor stage and consider them as the different time points for one patient. On the other hand, given the variety of genetically engineered mice models, one should carefully address their modeling targets and pick the model representing the corresponding tumor microenvironment as closely as possible. 

For the ODE modeling, we proposed a simple mass-action reaction network. The promotion and inhibition of cell and cytokine production can simply be achieved through cell-cell, cell-cytokine, and cytokine-cytokine interactions and modeled via linear positive and negative feedback terms. However, in terms of cooperative reactions and in case reversibility and reactant-stationary assumptions during the initial transient period are satisfied, one can incorporate Michaelis-Menten kinetics and rates. To avoid blow-ups for cells that are not differentiated from a naive state, we used logistic growth models with capacity terms. This includes cells such as tumor cells, adipocytes, epithelial cells, mast cells, and such.

Parameter estimation is one of the most challenging steps of each mathematical model. It is tightly related to the available data, and each method is only practiced if its advantages outweigh its disadvantages. For example, the widely used Markov Chain Monte Carlo methods require a good understanding of the parameter distributions and can be very slow for large parameter spaces. The parameter estimation based on steady-state assumptions is powerful when we only have data at the initial and final stages of cancer. However, the assumptions made in this method require a careful uncertainty analysis. Finally, the least square optimization is strong, especially when time course data is available. But it depends highly on data and requires lots of trial and error to avoid under or overfitting. 

To be able to model the spatial interaction of cells and molecules in the microenvironment, we need to extend our kinetic ODE models to PDEs. A benefit of the workflow introduced here is that we already have the kinetics and corresponding parameters. The other benefit was its natural connection with the mechanical growth to construct a free boundary model. Intuitively, our model relates cell proliferation to mechanical growth by triggering advection when a threshold density is surpassed at a point in the domain. Another strength of these models is their flexibility to work with different mechanical models such as linear or nonlinear elasticity, fluid dynamics, and even porous medium. However, these models have a main disadvantage when adding terms such as chemotaxis. Since a full decoupling of the biological and mechanical problems cannot be attained, they must be solved simultaneously, which increases the computational burden. 

We introduced sensitivity analysis as means to both measure our model's robustness and find important biological control parameters. We applied a global sensitivity analysis for the ODE parameters, which is especially powerful when the parameter estimation is carried out using steady-state assumptions, because it considers the effect of perturbing the parameter of interest on the rest of the parameters. Given the algebraic dependence of parameters on each other and the assumptions used in the steady-state method, the global sensitivity analysis is a great approach for a more in-depth analysis. However, when dealing with PDE systems due to time and space dependence, direct methods can be extremely tedious and time-consuming. Therefore, we introduced the adjoint-based sensitivity analysis. The idea behind this method is to obviate the dependence of the sensitivity on the solution of the system. This way, we do not have to calculate the solution of the system for each parameter sensitivity analysis.

There are some limitations and challenges that are worth mentioning. For starters, further investigation into the dynamics and the parameter space is challenging for large systems. Perhaps simplifying the model by comparing the time scales of reaching the steady-state values for each variable can be helpful. This way, we can replace the variables which reach their steady-state very fast with constants. Moreover, a good mathematical model requires a lot of maintenance and validation. This means close collaboration with biologists and clinicians and designing experiments that are specifically addressing the goals or flaws of the model. We hope such articles can motivate collaborations, experimental longitudinal studies, and better data for model validations which in turn leads to more reliable mathematical predictions.

%
%
%
\bibliographystyle{abbrv}
\bibliography{main}

\begin{thebibliography}{10}

\bibitem{cancer2013cancer}
D.~C. C. B. R. . J. M. A. . K. A. . P. T. . P. D. . W.~Y. 53 and T.~S. S. L. D.~A. 68.
\newblock The cancer genome atlas pan-cancer analysis project.
\newblock {\em Nature genetics}, 45(10):1113--1120, 2013.

\bibitem{Aronow2022}
R.~A. Aronow, S.~Akbarinejad, T.~Le, S.~Su, and L.~Shahriyari.
\newblock Tumordecon: A digital cytometry software.
\newblock {\em SoftwareX}, 18:101072, 6 2022.

\bibitem{attalla2021insights}
S.~Attalla, T.~Taifour, T.~Bui, and W.~Muller.
\newblock Insights from transgenic mouse models of pymt-induced breast cancer: recapitulating human breast cancer progression in vivo.
\newblock {\em Oncogene}, 40(3):475--491, 2021.

\bibitem{brawley202150}
O.~W. Brawley and P.~Goldberg.
\newblock The 50 years' war: The history and outcomes of the national cancer act of 1971.
\newblock {\em Cancer}, 127(24):4534--4540, 2021.

\bibitem{cai2017transcriptomic}
Y.~Cai, R.~Nogales-Cadenas, Q.~Zhang, J.-R. Lin, W.~Zhang, K.~O’Brien, C.~Montagna, and Z.~D. Zhang.
\newblock Transcriptomic dynamics of breast cancer progression in the mmtv-pymt mouse model.
\newblock {\em Bmc Genomics}, 18(1):1--14, 2017.

\bibitem{international2010international}
D.~coordination~centre Kasprzyk (Leader) Arek 1 Stein (Leader) Lincoln D. 1 Zhang Junjun 1 Haider Syed A. 98 Wang Jianxin 1 Yung Christina K. 1 Cross Anthony 1 Liang Yong 1 Gnaneshan Saravanamuttu 1 Guberman Jonathan 1 Hsu Jack~1 et~al.
\newblock International network of cancer genome projects.
\newblock {\em Nature}, 464(7291):993--998, 2010.

\bibitem{de1979kinetics}
A.~De~Lean and D.~Rodbard.
\newblock Kinetics of cooperative binding.
\newblock {\em General Principles and Procedures}, pages 143--192, 1979.

\bibitem{domschke2014mathematical}
P.~Domschke, D.~Trucu, A.~Gerisch, and M.~A. Chaplain.
\newblock Mathematical modelling of cancer invasion: implications of cell adhesion variability for tumour infiltrative growth patterns.
\newblock {\em Journal of theoretical biology}, 361:41--60, 2014.

\bibitem{edgar2002gene}
R.~Edgar, M.~Domrachev, and A.~E. Lash.
\newblock Gene expression omnibus: Ncbi gene expression and hybridization array data repository.
\newblock {\em Nucleic acids research}, 30(1):207--210, 2002.

\bibitem{franks2003modelling}
S.~Franks, H.~Byrne, J.~King, J.~Underwood, and C.~Lewis.
\newblock Modelling the early growth of ductal carcinoma in situ of the breast.
\newblock {\em Journal of mathematical biology}, 47:424--452, 2003.

\bibitem{gerstner1998numerical}
T.~Gerstner and M.~Griebel.
\newblock Numerical integration using sparse grids.
\newblock {\em Numerical algorithms}, 18(3-4):209, 1998.

\bibitem{gompertz1825xxiv}
B.~Gompertz.
\newblock Xxiv. on the nature of the function expressive of the law of human mortality, and on a new mode of determining the value of life contingencies. in a letter to francis baily, esq. frs \&c.
\newblock {\em Philosophical transactions of the Royal Society of London}, (115):513--583, 1825.

\bibitem{hao2014ldl}
W.~Hao and A.~Friedman.
\newblock The ldl-hdl profile determines the risk of atherosclerosis: a mathematical model.
\newblock {\em PloS one}, 9(3):e90497, 2014.

\bibitem{horn1972general}
F.~Horn and R.~Jackson.
\newblock General mass action kinetics.
\newblock {\em Archive for rational mechanics and analysis}, 47:81--116, 1972.

\bibitem{hu2022bio}
Y.~Hu, N.~Mohammad~Mirzaei, and L.~Shahriyari.
\newblock Bio-mechanical model of osteosarcoma tumor microenvironment: A porous media approach.
\newblock {\em Cancers}, 14(24):6143, 2022.

\bibitem{hughes2012quantitative}
S.~K. Hughes-Alford and D.~A. Lauffenburger.
\newblock Quantitative analysis of gradient sensing: towards building predictive models of chemotaxis in cancer.
\newblock {\em Current opinion in cell biology}, 24(2):284--291, 2012.

\bibitem{kirshtein2020data}
A.~Kirshtein, S.~Akbarinejad, W.~Hao, T.~Le, S.~Su, R.~A. Aronow, and L.~Shahriyari.
\newblock Data driven mathematical model of colon cancer progression.
\newblock {\em Journal of Clinical Medicine}, 9(12):3947, 2020.

\bibitem{kronik2008improving}
N.~Kronik, Y.~Kogan, V.~Vainstein, and Z.~Agur.
\newblock Improving alloreactive ctl immunotherapy for malignant gliomas using a simulation model of their interactive dynamics.
\newblock {\em Cancer Immunology, Immunotherapy}, 57:425--439, 2008.

\bibitem{Le2020}
T.~Le, R.~A. Aronow, A.~Kirshtein, and L.~Shahriyari.
\newblock A review of digital cytometry methods: estimating the relative abundance of cell types in a bulk of cells.
\newblock {\em Briefings in bioinformatics}, 10 2020.

\bibitem{le2021data}
T.~Le, S.~Su, A.~Kirshtein, and L.~Shahriyari.
\newblock Data-driven mathematical model of osteosarcoma.
\newblock {\em Cancers}, 13(10):2367, 2021.

\bibitem{lewis2021spatial}
S.~M. Lewis, M.-L. Asselin-Labat, Q.~Nguyen, J.~Berthelet, X.~Tan, V.~C. Wimmer, D.~Merino, K.~L. Rogers, and S.~H. Naik.
\newblock Spatial omics and multiplexed imaging to explore cancer biology.
\newblock {\em Nature methods}, 18(9):997--1012, 2021.

\bibitem{li2004adjoint}
S.~Li and L.~Petzold.
\newblock Adjoint sensitivity analysis for time-dependent partial differential equations with adaptive mesh refinement.
\newblock {\em Journal of Computational Physics}, 198(1):310--325, 2004.

\bibitem{mirzaei2023investigating}
N.~M. Mirzaei, W.~Hao, and L.~Shahriyari.
\newblock Investigating the spatial interaction of immune cells in colon cancer.
\newblock {\em iScience}, 2023.

\bibitem{mitra2019parameter}
E.~D. Mitra and W.~S. Hlavacek.
\newblock Parameter estimation and uncertainty quantification for systems biology models.
\newblock {\em Current opinion in systems biology}, 18:9--18, 2019.

\bibitem{mohammad2022investigating}
N.~Mohammad~Mirzaei, N.~Changizi, A.~Asadpoure, S.~Su, D.~Sofia, Z.~Tatarova, I.~K. Zervantonakis, Y.~H. Chang, and L.~Shahriyari.
\newblock Investigating key cell types and molecules dynamics in pymt mice model of breast cancer through a mathematical model.
\newblock {\em PLoS computational biology}, 18(3):e1009953, 2022.

\bibitem{mohammad2021mathematical}
N.~Mohammad~Mirzaei, S.~Su, D.~Sofia, M.~Hegarty, M.~H. Abdel-Rahman, A.~Asadpoure, C.~M. Cebulla, Y.~H. Chang, W.~Hao, P.~R. Jackson, et~al.
\newblock A mathematical model of breast tumor progression based on immune infiltration.
\newblock {\em Journal of Personalized Medicine}, 11(10):1031, 2021.

\bibitem{mohammad2022pde}
N.~Mohammad~Mirzaei, Z.~Tatarova, W.~Hao, N.~Changizi, A.~Asadpoure, I.~K. Zervantonakis, Y.~Hu, Y.~H. Chang, and L.~Shahriyari.
\newblock A pde model of breast tumor progression in mmtv-pymt mice.
\newblock {\em Journal of Personalized Medicine}, 12(5):807, 2022.

\bibitem{mohammad2020integrated}
N.~Mohammad~Mirzaei, W.~S. Weintraub, and P.-W. Fok.
\newblock An integrated approach to simulating the vulnerable atherosclerotic plaque.
\newblock {\em American Journal of Physiology-Heart and Circulatory Physiology}, 319(4):H835--H846, 2020.

\bibitem{newman2019determining}
A.~M. Newman, C.~B. Steen, C.~L. Liu, A.~J. Gentles, A.~A. Chaudhuri, F.~Scherer, M.~S. Khodadoust, M.~S. Esfahani, B.~A. Luca, D.~Steiner, et~al.
\newblock Determining cell type abundance and expression from bulk tissues with digital cytometry.
\newblock {\em Nature biotechnology}, 37(7):773--782, 2019.

\bibitem{painter2019mathematical}
K.~J. Painter.
\newblock Mathematical models for chemotaxis and their applications in self-organisation phenomena.
\newblock {\em Journal of theoretical biology}, 481:162--182, 2019.

\bibitem{paterson2020mathematical}
C.~Paterson, H.~Clevers, and I.~Bozic.
\newblock Mathematical model of colorectal cancer initiation.
\newblock {\em Proceedings of the National Academy of Sciences}, 117(34):20681--20688, 2020.

\bibitem{rens2020cell}
E.~G. Rens and R.~M. Merks.
\newblock Cell shape and durotaxis explained from cell-extracellular matrix forces and focal adhesion dynamics.
\newblock {\em Iscience}, 23(9):101488, 2020.

\bibitem{rodriguez1994stress}
E.~K. Rodriguez, A.~Hoger, and A.~D. McCulloch.
\newblock Stress-dependent finite growth in soft elastic tissues.
\newblock {\em Journal of biomechanics}, 27(4):455--467, 1994.

\bibitem{schnell2014validity}
S.~Schnell.
\newblock Validity of the michaelis--menten equation--steady-state or reactant stationary assumption: that is the question.
\newblock {\em The FEBS journal}, 281(2):464--472, 2014.

\bibitem{shangerganesh2019finite}
L.~Shangerganesh, N.~Nyamoradi, G.~Sathishkumar, and S.~Karthikeyan.
\newblock Finite-time blow-up of solutions to a cancer invasion mathematical model with haptotaxis effects.
\newblock {\em Computers \& Mathematics with Applications}, 77(8):2242--2254, 2019.

\bibitem{siegel2023cancer}
R.~L. Siegel, K.~D. Miller, N.~S. Wagle, and A.~Jemal.
\newblock Cancer statistics, 2023.
\newblock {\em CA: a cancer journal for clinicians}, 73(1):17--48, 2023.

\bibitem{sofia2022patient}
D.~Sofia, N.~Mohammad~Mirzaei, and L.~Shahriyari.
\newblock Patient-specific mathematical model of the clear cell renal cell carcinoma microenvironment.
\newblock {\em Journal of Personalized Medicine}, 12(10):1681, 2022.

\bibitem{Stahlberg2022}
E.~A. Stahlberg, M.~Abdel-Rahman, B.~Aguilar, A.~Asadpoure, R.~A. Beckman, L.~L. Borkon, J.~N. Bryan, C.~M. Cebulla, Y.~H. Chang, A.~Chatterjee, J.~Deng, S.~Dolatshahi, O.~Gevaert, E.~J. Greenspan, W.~Hao, T.~Hernandez-Boussard, P.~R. Jackson, M.~Kuijjer, A.~Lee, P.~Macklin, S.~Madhavan, M.~D. McCoy, N.~M. Mirzaei, T.~Razzaghi, H.~L. Rocha, L.~Shahriyari, I.~Shmulevich, D.~G. Stover, Y.~Sun, T.~Syeda-Mahmood, J.~Wang, Q.~Wang, and I.~Zervantonakis.
\newblock Exploring approaches for predictive cancer patient digital twins: Opportunities for collaboration and innovation.
\newblock {\em Frontiers in Digital Health}, 4, 10 2022.

\bibitem{stroberg2016estimation}
W.~Stroberg and S.~Schnell.
\newblock On the estimation errors of km and v from time-course experiments using the michaelis--menten equation.
\newblock {\em Biophysical chemistry}, 219:17--27, 2016.

\bibitem{trucu2013multiscale}
D.~Trucu, P.~Lin, M.~A. Chaplain, and Y.~Wang.
\newblock A multiscale moving boundary model arising in cancer invasion.
\newblock {\em Multiscale Modeling \& Simulation}, 11(1):309--335, 2013.

\bibitem{tsoularis2002analysis}
A.~Tsoularis and J.~Wallace.
\newblock Analysis of logistic growth models.
\newblock {\em Mathematical biosciences}, 179(1):21--55, 2002.

\bibitem{valderrama2019mcmc}
G.~I. Valderrama-Baham{\'o}ndez and H.~Fr{\"o}hlich.
\newblock Mcmc techniques for parameter estimation of ode based models in systems biology.
\newblock {\em Frontiers in Applied Mathematics and Statistics}, 5:55, 2019.

\bibitem{wang2022dynamics}
H.~Wang, C.~Zhao, C.~A. Santa-Maria, L.~A. Emens, and A.~S. Popel.
\newblock Dynamics of tumor-associated macrophages in a quantitative systems pharmacology model of immunotherapy in triple-negative breast cancer.
\newblock {\em iScience}, 25(8):104702, 2022.

\bibitem{wang2020review}
Y.~Wang.
\newblock A review on the qualitative behavior of solutions in some chemotaxis--haptotaxis models of cancer invasion.
\newblock {\em Mathematics}, 8(9):1464, 2020.

\bibitem{zi2011sensitivity}
Z.~Zi.
\newblock Sensitivity analysis approaches applied to systems biology models.
\newblock {\em IET systems biology}, 5(6):336--346, 2011.

\bibitem{zimmermann2004genevestigator}
P.~Zimmermann, M.~Hirsch-Hoffmann, L.~Hennig, and W.~Gruissem.
\newblock Genevestigator. arabidopsis microarray database and analysis toolbox.
\newblock {\em Plant physiology}, 136(1):2621--2632, 2004.

\end{thebibliography}

\end{document}